# Fabrication and Characterization of Short Josephson Junctions with Stepped Ferromagnetic Barrier

Uthayasankaran Peralagu and Martin Weides

Institute for Solid State Research, Research Centre Juelich, 52425 Juelich, Germany
e-mail: m.weides@fz-juelich.de

***Abstract*—We present novel low-$T_c$ superconductor-insulator-ferromagnet-superconductor (SIFS) Josephson junctions with planar and stepped ferromagnetic interlayer. We optimized the fabrication process to set a step in the ferromagnetic layer thickness. Depending on the thickness of the ferromagnetic layer the ground state of the SIFS junction has a phase drop of either 0 or $\pi$. So-called 0–$\pi$ Josephson junctions, in which 0 and $\pi$ ground states compete with each other, were obtained. These stepped junctions may have a double degenerate ground state, corresponding to a vortex of supercurrent circulating clock- or counterclockwise and creating a magnetic flux which carries a fraction of the magnetic flux quantum $\Phi_0$. Here, we limit the presentation to static properties of short junctions.**

## I. INTRODUCTION

Superconductivity (S) and ferromagnetism (F) are two competing phenomena due to their unequal time symmetry: ferromagnetic order breaks the time-reversal symmetry, whereas conventional superconductivity relies on the pairing of time-reversed states. It turns out that the combination of both leads to rich and interesting physics. One particular example – the phase oscillations of the superconducting Ginzburg-Landau order parameter inside the ferromagnet [1] –plays a major role for the devices discussed in this work. If the thickness $d_F$ of the ferromagnetic barrier in SIFS-type (I: insulating tunnel barrier) Josephson junction (JJ) is on the order of one half this oscillation wave length, the order parameter changes its sign, *i.e.,* shifts its phase by π while crossing the ferromagnet. In this case the critical current $I_c$ (and critical current density $j_c$) turns out to be negative and the current-phase relation reads $I = I_c \sin(\varphi) = |I_c| \sin(\varphi+\pi)$ with $I_c < 0$. Such a JJ is called "π JJ" because it has $\varphi = \pi$ in the ground state. Conventional JJs are called "0 JJ" because they have a current-phase relation of $I = I_c \sin(\varphi)$ with $I_c > 0$ and the ground state phase $\varphi = 0$. The change in the sign of $I_c$ was shown to occur as a function of temperature [2] and of the ferromagnetic barrier thickness [3,4,5]. A quantitative model describing the behavior of $I_c$ and density of states as a function of parameters characterizing material properties of the S, F layers and the S/F interface transparency can be found in reference [6].

By using a ferromagnetic barrier with variable step-like thickness along the junction, we obtained a so-called 0-π Josephson junction [7], in which 0 and π ground states compete with each other. For well chosen thicknesses the 0 and π parts of the junction are perfectly symmetric, i.e. their absolute critical current densities $j_c^0$, $j_c^\pi$ are equal. In this case the degenerate ground state corresponds to a vortex of supercurrent circulating clockwise- or counterclockwise, thereby creating a magnetic flux which carries a fraction of the magnetic flux quantum $\Phi_0$ [8].

For 0–π junctions one needs 0 and π coupling in one junction, setting high demands on the fabrication process. The ideal 0–π JJ would have equally $j_c^0 = \left|j_c^\pi\right|$ and a 0–π phase boundary in its center to have a symmetric situation. Furthermore the junctions should be underdamped (SIFS structure) since low dissipation is necessary to study dynamics and eventually macroscopic quantum effects. The junctions should have a high $j_c$ (and hence small $\lambda_J \propto \sqrt{1/j_c}$) to reach the long JJ limit and to keep high $V_c = I_c R$ products, where $V_c$ is the characteristic voltage and $R$ the normal state resistance. Previous experimental works on 0–π JJs based on SFS technology [9,10] gave no information about $j_c^0$ and $j_c^\pi$. Hence, the ratio of asymmetry $\Delta = |j_c^\pi| / j_c^0$ was unknown and the Josephson penetration depth $\lambda_J$ could not be calculated for these samples. The first intentionally made symmetric 0–π tunnel JJ of SIFS type with a large $V_c$ was realized by the authors [7], making direct transport measurements of $I_c(H)$ and calculation of the ground state with spontaneous flux feasible.

New superconducting spintronic devices such as FSF spin valves and SIFS Josephson junctions with 0, π and 0-π coupling have gained considerable interest in recent years because they show a number of interesting properties for future classical and quantum computing devices.

## II. DEPOSITION AND PATTERNING

The fabrication process [11] is based on Nb/Al-$Al_2O_3$/Cu/NiCu/Nb stacks, see Figure 1, deposited by dc magnetron sputtering on thermally oxidized 4-inch Si substrate. The 160 nm thick Nb bottom electrode, made up by four 40 nm Nb layers, each separated by 2.4 nm Al layers to reduce roughness, was covered by a 5 nm thick Al layer and oxidized for 30 min at RT and $10^{-2}$ to 1 Pa residual oxygen pressure. To obtain many structures with different F-layer thickness in one fabrication run, we deposit a wedge-shaped F-layer (i.e. $Ni_{0.6}Cu_{0.4}$) alloy in order to minimize inevitable run-to-run variations. The 2.6 nm thin Cu interlayer provided smooth growth of the NiCu layer. The multilayer stack was covered with a 40 nm Nb cap layer.

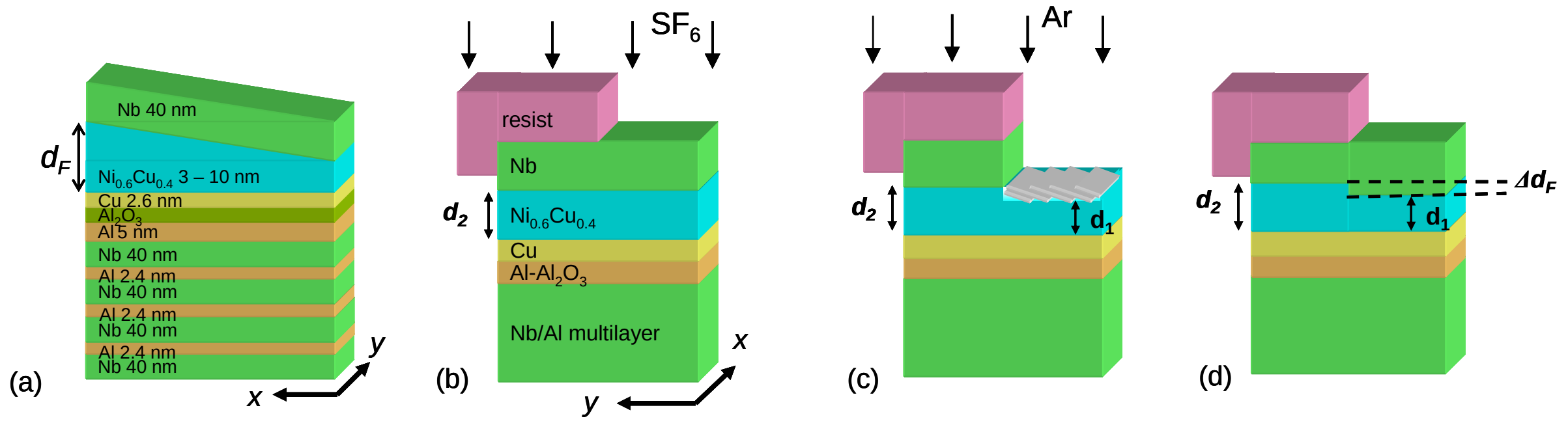

**Fig. 1.** (a) Cross-section of SIFS multilayer, (b) reactive etching of Nb with $SF_6$ down to the NiCu layer, (c) ion-etching of NiCu to set 0 coupling and (d) in situ deposition of the cap Nb layer (d). More details can be found in [12].

The stepped junctions were patterned using a four-level optical photolithographic mask procedure including $SF_6$ reactive etching and Ar ion-beam milling [12]. The step in F-layer was patterned after lift-off. The junction was partly protected with photoresist to define the step location in the F-layer, followed by i) selective $SF_6$ reactive etching of the Nb, ii) Ar ion-etching of the NiCu by $\Delta d_F$ and iii) subsequent in situ deposition of Nb. The lithographic accuracy is on the order of 1 μm. The insulating layer between top and bottom electrode is self-aligned by ion-beam etching below the Al layer and anodic oxidation of the bottom Nb electrode. Finally the top wiring was deposited.

We increased the reproducibility of samples, as depicted in Figure 2, by optimizing: i) the interface quality by employing Al interlayers to smoothen the top Al-$Al_2O_3$ layer, ii) the control of step formation by adjusting the etching stage and iii) the F-layer deposition parameters. These modifications may contribute to the changes in nominal F-layer thickness, at which the 0 to π phase transition point occurs, compared to previous publications [7,12, 13].

Three samples (named 1st, 2nd and 3rd) were fabricated. The partial oxygen pressure for tunnel barrier formation decreased systematically from sample to sample to get a thinner tunnel barrier with larger critical current densities. The etching times for the step formation were adjusted to obtain symmetric 0-π JJs with the highest possible $j_c^0$ and $j_c^\pi$.

For each sample various junctions on the wafer were placed within a narrow row perpendicular to the gradient in the F-layer thickness ($x$-axis) and were replicated along this gradient. One row contained a triplet of junctions: reference JJs with the uniform F-layer thickness $d_1$ (uniformly etched) and $d_2$ (non-etched) and a JJ with step $\Delta d_F$ in the F-layer thickness from $d_1$ to $d_2$. In Figure 2 we plot the F-layer thickness dependence of the critical current density in SIFS junctions (filled symbols: non-etched JJs, open symbols: etched JJs) for three samples as function of $d_F$. At $d_F \approx 6.8$ nm (for non-etched JJs) the critical current is nearly vanishing as the order parameter is zero when changing from 0 to π. For the other two samples this region was not traced out, as the their $j_c(d_F)$ curve just became lifted to larger $j_c$ values [5].

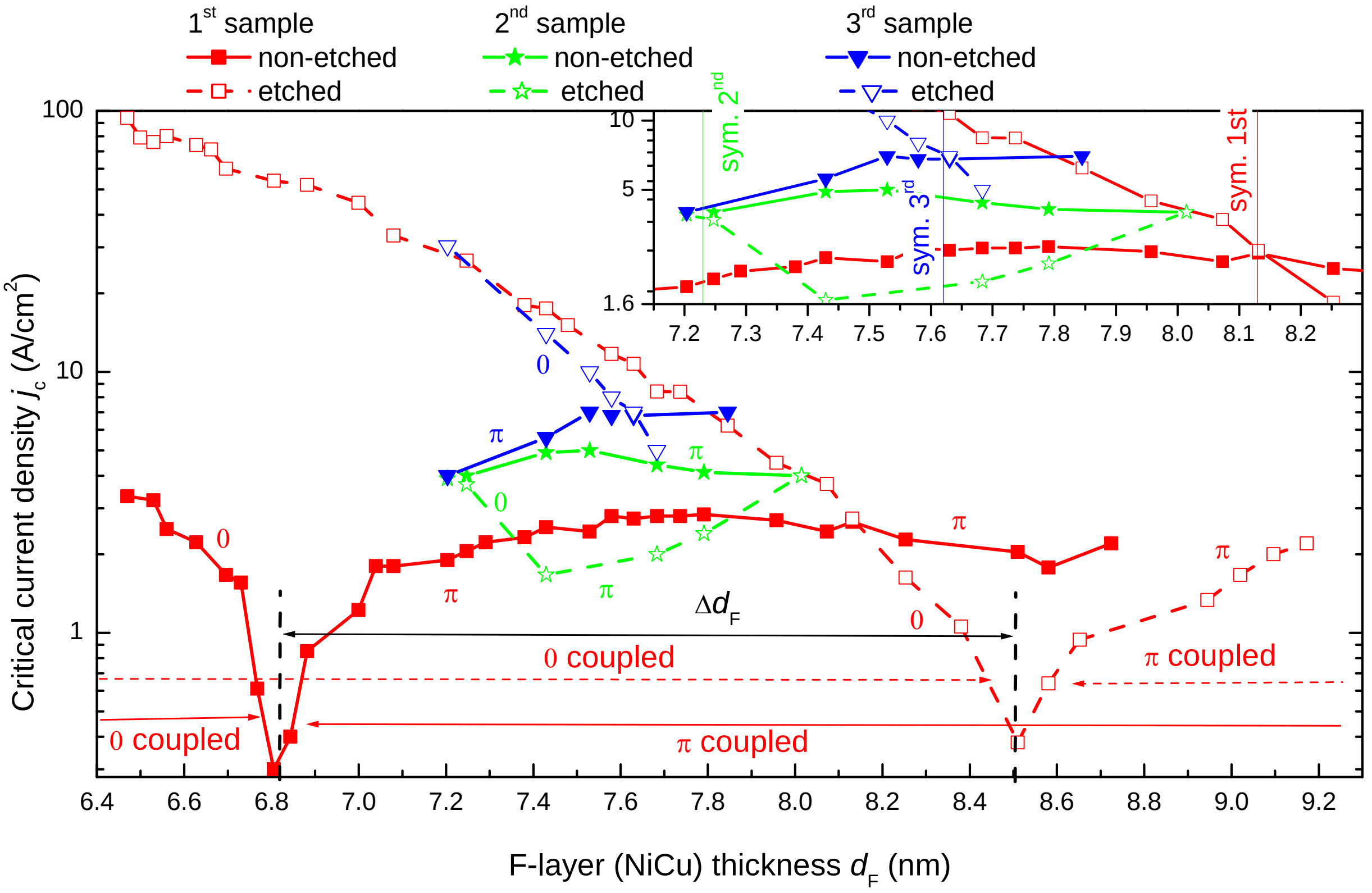

**Fig. 2.** Dependence of critical current density $j_c$ for SIFS reference JJs that were not etched (filled symbols, $d_2$) and etched uniformly (open symbols, $d_1$) on the thickness of the F-layer $d_F$. The lines are guides for the eyes. The tunnel barrier thickness decreased from sample 1st to 3rd. The vertical lines in the inset mark $d_F$ values where $j_c^0 = \left| j_c^\pi \right|$, *i.e.*, where symmetric 0-π JJs were obtained.

We estimate the etched-away F-layer thickness as $\Delta d_F \approx 1.7$ nm for the 1st sample by comparing the critical current densities $j_c$ of non-etched JJs with the $j_c(d_F)$ data for the etched samples. For 2nd and 3rd sample $\Delta d_F$ it is even smaller. The good control over the step height is shown by the very low F-layer etching rate (0.0215 nm/sec).

## III. CHARACTERIZATION

Studying the current-voltage and $I_c(H)$ characteristics for the planar reference 0 and π JJs is the starting point for estimation of the ground state of a stepped JJ. From these characteristics one can calculate important parameters such as the critical current densities $j_c^0$, $j_c^\pi$, the Josephson penetration depths $\lambda_J^0$, $\lambda_J^\pi$ and the ratio of asymmetry $\Delta = |j_c^\pi| / j_c^0$.

We choose triplets of junctions which have the thickness $d_2$ and $j_c(d_2) < 0$ (π junction) before etching and the thickness $d_1 = d_2 - \Delta d_F$ and $j_c(d_1) \approx -j_c(d_2)$ (0 junction) after etching. The vertical lines in the inset of Figure 2 depict the location of triplets for the three samples.

For simplicity in this article we will concentrate on the static properties of short 0-π JJs, namely the magnetic diffraction pattern which can be solved analytically. The dynamic properties of short and intermediate 0-π JJs are described in a separate publication [14]. The junctions are taken from the first sample ($d_1$=6.45 nm, $d_2$=8.15 nm). They have a length of 100 μm and a width of 50 μm. The 0 JJ has $d_F = d_1$, the π JJ has $d_F = d_2$ and the 0–π JJs with stepped F-layer $d_F = d_1$, $d_2$ in each half. The magnetic field is applied perpendicular to the long axis and parallel to the step in the F-layer (i.e. || $x$-axis).

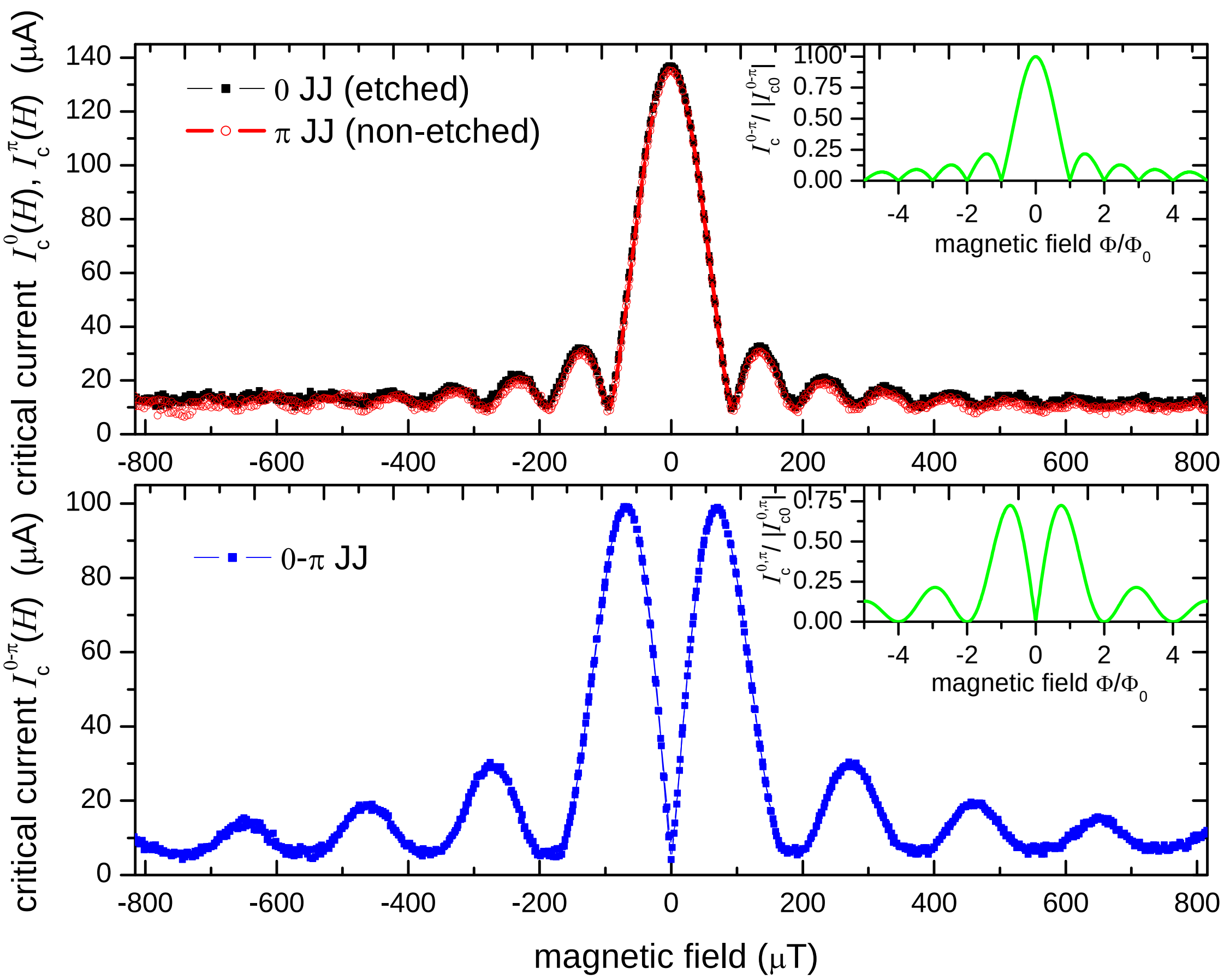

**Fig. 3.** $I_c(H)$ of reference 0, π and stepped 0-π JJs. The insets show analytical calculated $I_c(H)$ pattern for short JJs limit. The 0-π JJ is symmetric because $j_c^0 = |j_c^\pi|$. Measurements were done at 4.2 K.

The magnetic diffraction pattern $I_c(H)$ of the 0–π JJ and the 0 and π reference JJs are plotted in Figure 3 together with the calculated pattern for a 0–π JJs [15]

$$I_c^{0-\pi} = \left|I_{c0}^{0-\pi}\right| \sin^2\left(\frac{\pi}{2}\frac{\Phi}{\Phi_0}\right) / \left|\frac{\pi}{2}\frac{\Phi}{\Phi_0}\right|$$

and for 0 or π JJs

$$I_c^{0,\pi} = \left|I_{c0}^{0,\pi}\right| \sin\left(\pi\frac{\Phi}{\Phi_0}\right) / \left(\pi\frac{\Phi}{\Phi_0}\right)$$

as depicted in the insets. Φ is the magnetic flux induced by $H$ on the cross-section of JJs. For a 0–π JJ the $I_c(0)$ should vanish, the maxima $I_c$'s are symmetric and are about $0.72 \cdot I_{c0}^{0,\pi}$. Due to a small net magnetization of the F-layers the $I_c(H)$ of references junctions were sometimes slightly shifted along the $H$ axis. The magnetic field dependence of the planar reference junctions $I_c^0(H)$ and $I_c^\pi(H)$ look like perfect Fraunhofer pattern. One can see that both magnetic diffraction pattern almost coincide, having the form of a symmetric Fraunhofer pattern with the critical currents $I_c^{0,\pi}(0) \approx 136\ \mu\text{A}$ and the same oscillation period $\mu_0 H_{c1} \approx 93$ μT. The 0–π JJ had a magnetic field dependence $I_c^{0-\pi}(H)$ with a clear minimum near zero magnetic field ($I_c^{0-\pi}(0) \approx 4\ \mu\text{A}$) and almost no asymmetry ($\Delta \approx 1$). The critical currents at the left and right maxima ($99.2$ and $98.9\ \mu\text{A}$) differ by less than 2%, and are $\approx 0.72 \cdot I_{c0}^{0,\pi}$, as expected from theory. The voltage criteria of 1 μV for the determination of $I_c$ accounts for the offset along current axis for all minima. The oscillation period is $\mu_0 H_{c1} \approx 184$ μT, nearly the double of the planar junctions, as expected from theory.

The 0-π JJ is symmetric, because the 0 to π phase step is centred and the critical current densities in both parts have equal amplitude ($j_c^0 = |j_c^\pi|$).The junction length is estimated as $\approx 0.44\lambda_J$, thus being clearly within the short JJ limit. Therefore in ground state the spontaneous flux in the junction is very small $\Phi = \pm \frac{\Phi_0}{8\pi} \cdot 0.44^2$, *i.e.* $|\Phi| \approx 0.0077 \cdot \Phi_0$.

## IV. OUTLOOK

As an outlook, the fabrication of stepped Josephson junctions allows tailoring of ferromagnetic interlayer thickness to achieve, for example, so-called 0-π coupled junctions where the physics of fractional vortices can be studied. The presented SIFS technology allows us to construct 0, π and 0–π JJs with comparable $j_c^0$ and $|j_c^\pi|$ in a single fabrication run. Such JJs may be used to build classical and quantum devices such as oscillators, memory cells, superconducting flux qubits with a π junction [16], or fractional flux quantum based qubits [17]. Josephson junctions with varying $j_c$ and planar phase (0 or π) could be used for devices with a specially shaped $I_c(H)$ pattern [18], toy systems for flux pinning, or tunable superconducting resonators. The 0-π JJs with a stepped F-layer and low-$T_c$ superconductor electrodes offer great flexibility for the integration of these devices, as they show advantages over the existing 0-π junctions based on d-wave superconductors [19,20] or current injectors [21] such as low dissipation of plasma oscillations, no restrictions in topology, no additional bias electrodes, and easy integration into the mature Nb/Al-$Al_2O_3$/Nb technology.

The technology for step formation can be applied to other metallic multilayer systems such as magneto-resistance devices (GMR/TMR elements) where a local variation of magnetic properties may enhance their functionality.

## ACKNOWLEDGMENT

We thank H. Kohlstedt and E. Goldobin for useful discussions. M. W. thanks DFG (project WE 4359/1-1) for financial support.